\documentclass[sigconf]{acmart}

\renewcommand\footnotetextcopyrightpermission[1]{}
\setcopyright{none}
\acmConference{}{}{}{}
\acmBooktitle{}

\usepackage{booktabs}
\usepackage{algorithm}
\usepackage{algpseudocode}
\usepackage[most]{tcolorbox}
\usepackage{tikz}
\usetikzlibrary{arrows.meta, positioning, fit, calc, backgrounds, shapes.geometric, matrix}

\begin{document}

\title{MoSAIC-X: Architecture-Aware Scheduling of Tiled Neural-Network Kernels on GPUs via Solver-Calibrated Priority Search}

\author{Shiva Ahir$^{1}$, Hashmath Shaik$^{1}$, Alex Doboli$^{1}$, Nadeem Yaseen$^{2}$, Simona Doboli$^{3}$}
\affiliation{%
  \institution{%
    $^{1}$Department of Electrical \& Computer Engineering, Stony Brook University, New York, United States of America\\
    $^{2}$Barcelona Supercomputing Center, Barcelona, Catalonia, Spain\\
    $^{3}$Department of Computer Science, Hofstra University, New York, United States of America
  }
  \country{}
}
\email{{\{shiva.ahir,hashmath.shaik,alex.doboli\}}@stonybrook.edu}
\email{nadeem.yaseen@bsc.es, simona.doboli@hofstra.edu}

\begin{abstract}
  GPU-accelerated neural network training decomposes computation into tiled matrix operations that form large task DAGs with thousands of interdependent kernels. Scheduling these DAGs effectively requires balancing task priorities with GPU memory hierarchy constraints; yet classical heuristics use fixed rules that ignore architecture, exact solvers cannot scale, and neural schedulers sacrifice interpretability. This paper presents MoSAIC-X, a framework that learns compact, interpretable scheduling heuristics by calibrating against exact solver solutions. MoSAIC-X represents scheduling priority as a linear combination of five structural DAG features (upward rank, depth, fanout, indegree, and communication cost) and uses a staged stochastic search to find weights that minimize makespan, with the CP-SAT constraint solver providing optimal makespans as calibration references. Across PolyBench/MachSuite kernels, synthetic graph families, and Transformer training DAGs with up to 3{,}333 tasks, MoSAIC-X achieves the best or tied-best makespan on seven of eight community benchmarks, stays within 0.5\% of verified optimal solutions, and reduces makespan by up to 12\% over HEFT on communication-heavy workloads. Hardware validation on an NVIDIA RTX 500 Ada GPU confirms a measured $1.015\times$ speedup on a 901-task configuration. Compared to the 417-parameter neural scheduler DeepSoCS, MoSAIC-X achieves superior scheduling quality using just five interpretable parameters.
\end{abstract}


\maketitle

\section{Introduction}

Efficient task scheduling for Directed Acyclic Graphs (DAGs) is a fundamental problem in heterogeneous computing, spanning system-on-chip (SoC) architectures, GPU-accelerated training, and high-level synthesis~\cite{ref22, ref23, ref24}. The goal is to assign tasks to processing elements and determine their execution order so that the makespan is minimized while satisfying precedence and resource constraints.

The scheduling problem is NP-hard~\cite{ref22}, motivating three broad families of solutions. Heuristic algorithms such as HEFT and CPOP~\cite{ref01}, PEFT~\cite{ref02}, DLS~\cite{ref03}, and FDS~\cite{ref04} remain widely used due to their low computational cost~\cite{ref23, ref24}, but they rely on fixed priority rules that cannot adapt to diverse workloads. Constraint programming solvers like CP-SAT~\cite{ref07} achieve provably optimal schedules but scale poorly beyond moderate instance sizes. Learning-based approaches~\cite{ref17, ref18, ref19, ref20, ref21}, including neural schedulers such as DeepSoCS~\cite{ref09}, adapt to varying workloads but typically sacrifice interpretability, require hundreds of parameters, and need extensive per-graph training.

GPU-accelerated neural network training introduces a scheduling challenge that none of these families adequately addresses. Tiled matrix operations decompose training into hundreds to thousands of fine-grained tasks with structured dependencies (fork, fan-in, chain patterns) and data movement costs that depend on the GPU's memory hierarchy: shared memory capacity determines tile size, L2 cache residency determines whether inter-tile communication hits fast cache or slow DRAM, and scheduling order directly controls which data remains cache-resident. Classical heuristics like HEFT ignore this hierarchy entirely, treating all communication as uniform-cost edges. Exact solvers cannot scale to the resulting DAG sizes. Neural schedulers require retraining for each new workload and produce opaque policies. What is missing is a scheduling method that is \emph{simultaneously} near-optimal, interpretable, architecture-aware, and fast enough for online deployment.

In our prior work~\cite{ref05}, we introduced MoSAIC, a framework that searches for compact priority functions calibrated against CP-SAT optimal solutions over structural DAG features. MoSAIC demonstrated that a small set of learned weights can recover near-optimal scheduling behavior on synthetic and standard benchmarks (PolyBench~\cite{ref11}, MachSuite~\cite{ref12}). However, MoSAIC was evaluated only on general-purpose DAGs without memory hierarchy modeling or real GPU deployment.

This paper introduces MoSAIC-X, which extends MoSAIC to GPU scheduling of tiled neural network kernels. The core insight is that a five-parameter linear priority function over structural DAG features, when calibrated against exact solver solutions and combined with GPU memory hierarchy modeling, can produce near-optimal, fully interpretable schedules that transfer across workloads. We validate this on Transformer training DAGs with up to 1{,}024 tasks and matmul DAGs with up to 3{,}333 tasks, demonstrate real hardware speedup on NVIDIA RTX 500 Ada, and show that MoSAIC-X matches or outperforms baselines using orders of magnitude fewer parameters than neural alternatives.

Throughout this paper, ``on-chip'' refers to the \emph{target deployment context} (GPU SMs, shared memory, L2 cache); the priority search itself runs offline on a CPU.

\begin{figure*}[!htbp]
      \centering
      \resizebox{0.88\textwidth}{!}{%
      \begin{tikzpicture}[
          font=\scriptsize,
          >=Latex,
          every node/.style={align=center},
          box1/.style={draw, rounded corners=2pt, thick, minimum width=2.4cm, minimum height=0.8cm, fill=blue!14},
          box2/.style={draw, rounded corners=2pt, thick, minimum width=2.4cm, minimum height=0.8cm, fill=orange!18},
          box3/.style={draw, rounded corners=2pt, thick, minimum width=2.4cm, minimum height=0.8cm, fill=green!16},
          box4/.style={draw, rounded corners=2pt, thick, minimum width=2.6cm, minimum height=0.8cm, fill=purple!14},
          box5/.style={draw, rounded corners=2pt, thick, minimum width=2.6cm, minimum height=0.8cm, fill=red!14},
          aux/.style={draw, rounded corners=2pt, thick, minimum width=2.2cm, minimum height=0.7cm, fill=yellow!18},
          outbox/.style={draw, rounded corners=2pt, thick, minimum width=2.6cm, minimum height=0.8cm, fill=cyan!16},
          group/.style={draw, dashed, rounded corners=4pt, thick, inner sep=4pt},
          arr/.style={->, thick},
          fb/.style={->, thick, dashed}
      ]

      \node[box1] (dag) at (0,0) {\textbf{Benchmark DAG Corpus}\\STG, TGFF, PolyBench,\\MachSuite, Transformer};
      \node[box2] (parse) at (3.6,0) {\textbf{Feature Extraction}\\$\varphi(v) = [\text{rank}_u, \text{depth},$\\$\text{fanout}, \text{indeg}, \bar{c}_{\text{out}}]$};
      \node[box3] (cpsat) at (7.2,0) {\textbf{CP-SAT Solver}\\Optimal makespan\\$C_{\max}^*$ (reference)};
      \node[box4] (bo) at (11.0,0) {\textbf{Staged Stochastic Search}\\Recover $\theta^* \in \mathbb{R}^5$\\minimizing makespan};
      \node[box5] (kb) at (15.0,0) {\textbf{Learned Knowledge Base}\\$\{(G_i,\,\varphi_i,\,\theta_i^*)\}_{i=1}^N$};

      \node[aux] (gpu) at (1.8,-2.2) {\textbf{GPU Architecture}\\RTX 500 Ada: 16 SMs,\\48\,KB smem, 16\,MB L2};
      \node[aux] (l2) at (6.2,-2.2) {\textbf{L2 Cache Model}\\Locality-aware runtime\\adjustment per processor};
      \node[outbox] (transfer) at (11.0,-2.2) {\textbf{Zero-Shot Transfer}\\Apply $\theta^*$ from source\\DAGs to unseen targets};
      \node[outbox] (sched) at (15.6,-2.2) {\textbf{List Scheduling}\\Near-optimal schedule\\with 5-parameter rule};

      \draw[arr] (dag.east) -- (parse.west);
      \draw[arr] (parse.east) -- (cpsat.west);
      \draw[arr] (cpsat.east) -- (bo.west);
      \draw[arr] (bo.east) -- (kb.west);

      \draw[arr] (gpu.north) -- ++(0,0.4) -| ([xshift=-0.3cm]parse.south);
      \draw[arr] (gpu.east) -- (l2.west);
      \draw[arr] (l2.north) -- ++(0,0.4) -| ([xshift=-0.3cm]bo.south);
      \draw[arr] (kb.south) -- ++(0,-0.5) -| ([yshift=4pt]transfer.north);
      \draw[arr] (transfer.east) -- (sched.west);

      \draw[fb] (sched.north) -- ++(0,1.0) -| ([xshift=0.3cm]bo.south);

      \node[group, fit=(dag)(parse)(cpsat)(bo)(kb),
            label={[font=\bfseries\scriptsize]above:Offline: priority weight search calibrated against CP-SAT}] {};
      \node[group, fit=(gpu)(l2)(transfer)(sched),
            label={[font=\bfseries\scriptsize]below:Deployment: scheduling GPU workloads with learned priorities}] {};

      \end{tikzpicture}%
      }
      \vspace{-4pt}
      \caption{Overview of the MoSAIC-X framework. Top row: offline pipeline from benchmark DAGs through CP-SAT calibration to learned priority weights $\theta^* \in \mathbb{R}^5$. Bottom row: deployment on NVIDIA RTX 500 Ada with L2-cache-aware scheduling and zero-shot transfer to unseen DAGs. The CP-SAT solver provides optimal makespans as calibration references; the search itself directly minimizes the heuristic's makespan.}
      \label{fig:mosaic_framework}
  \end{figure*}
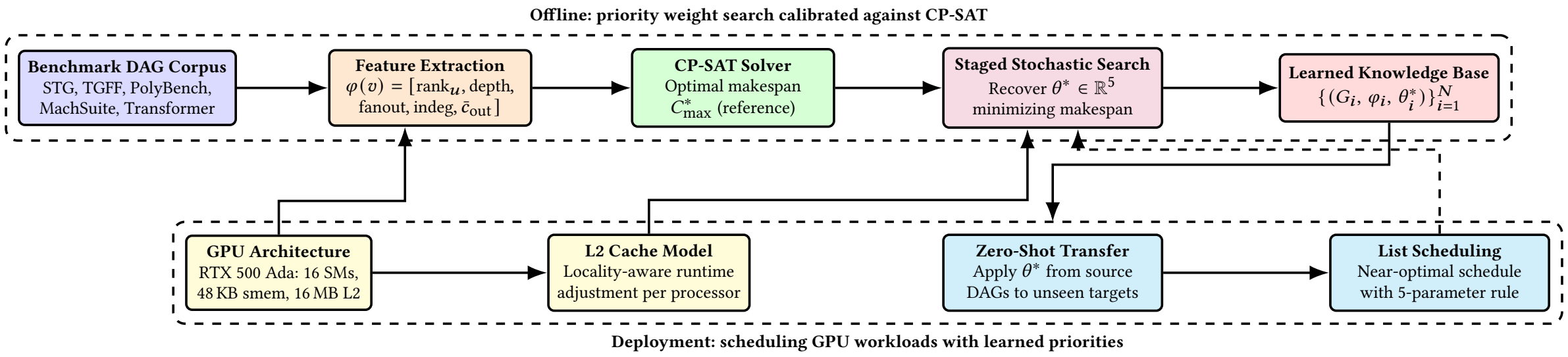

Beyond our prior work~\cite{ref05}, the key contributions are:
  \begin{enumerate}
      \item We show that tiled neural network training on GPUs can be effectively scheduled by a five-parameter priority function that respects shared memory and L2 cache constraints, achieving makespans within 0.5\% of proven optima on benchmarks where CP-SAT can certify optimality.
      \item We demonstrate that communication-heavy workloads benefit most from learned scheduling: MoSAIC-X reduces makespan by up to 12\% over HEFT when communication costs dominate, revealing that adaptive weighting of structural features captures scheduling opportunities that fixed heuristics miss.
      \item We validate on real GPU hardware (NVIDIA RTX 500 Ada), confirming that simulation-predicted improvements translate to measurable speedup on larger workloads.
      \item We show that a compact, interpretable model (5 parameters) can match or outperform a 417-parameter neural scheduler (DeepSoCS), suggesting that the scheduling problem's structure admits far simpler solutions than deep learning approaches assume.
  \end{enumerate}

The rest of the paper is organized as follows. Section~\ref{sec:related} surveys related work. Section~\ref{sec:method} presents the MoSAIC-X framework. Section~\ref{sec:experiments} describes the experimental setup and evaluation, followed by conclusions.

\section{Related Work}
\label{sec:related}

Task scheduling for DAGs on heterogeneous platforms has been studied extensively. We organize related work into three categories.

\paragraph{Heuristic schedulers.}
List-based heuristics remain the most widely used approach due to their low computational cost. HEFT and CPOP~\cite{ref01} rank tasks by upward or downward priority and assign each to the processor that minimizes finish time. PEFT~\cite{ref02} refines this with an optimistic cost table, while FDS~\cite{ref04} uses force-directed scheduling originally developed for high-level synthesis. DLS~\cite{ref03} combines static and dynamic priorities for interconnection-constrained architectures. These methods are fast and widely deployed~\cite{ref23, ref24}, and similar execution-aware optimization strategies have proven effective on resource-constrained embedded platforms~\cite{ref06}. However, they rely on fixed priority rules derived from a single structural property (typically upward rank) and cannot adapt their behavior to different graph topologies or exploit architecture-specific memory hierarchy effects~\cite{ref23}.

\paragraph{Exact and optimization-based methods.}
Constraint programming solvers such as CP-SAT~\cite{ref07} formulate scheduling as a constraint satisfaction problem and return provably optimal solutions. While valuable as calibration references, they scale poorly beyond moderate instance sizes (roughly a thousand tasks) and are impractical for the large DAGs arising in neural network training. The standard benchmark suites STG~\cite{ref14} and TGFF~\cite{ref15} provide synthetic task graphs used to evaluate both heuristic and exact approaches.

\paragraph{Learning-based schedulers.}
Recent work applies machine learning to scheduling, including deep reinforcement learning for resource management~\cite{ref17}, device placement optimization~\cite{ref18}, and neural combinatorial optimization~\cite{ref19, ref20, ref21}. DeepSoCS~\cite{ref09} combines graph neural network embeddings with deep RL for SoC scheduling, improving over classical heuristics but requiring hundreds of parameters and extensive per-graph training. Recent benchmark suites for heterogeneous SoC design~\cite{ref16} further highlight the need for scalable scheduling methods that can handle diverse hardware configurations. More broadly, research on network motifs~\cite{ref26} has shown that repeating subgraph patterns encode important structural information, yet motifs have not been directly leveraged for scheduling strategy design.

\paragraph{Our prior work.}
In Ahir et al.~\cite{ref05}, we introduced the MoSAIC framework, which searches for compact priority functions calibrated against CP-SAT optimal solutions over structural DAG features. That work demonstrated near-optimal scheduling on synthetic graphs and PolyBench/MachSuite kernels~\cite{ref11, ref12}, with zero-shot transfer across graph families. The broader theme of AI-assisted hardware design~\cite{ref13, ref10} motivates the present extension. MoSAIC-X adapts MoSAIC to GPU-based neural network training workloads with architecture-aware tiling, memory hierarchy modeling, Transformer-scale DAGs, real hardware validation, and comparison with neural scheduling baselines.

 \section{MoSAIC-X Framework Methodology}
 \label{sec:method}

  \subsection{Overview of the Methodology}

  The central idea is to replace fixed, hand-designed scheduling rules with a learnable priority function. Given a task graph with heterogeneous execution costs and communication weights mapped onto a GPU architecture, each task is described by a compact feature vector capturing its position and connectivity within the DAG. The scheduling priority is then a simple weighted combination of these features, where the weights are shared across all tasks in a graph.

  To find good weights, MoSAIC-X uses a staged stochastic search procedure. The CP-SAT constraint solver~\cite{ref07} provides provably optimal makespans on tractable DAGs, which serve as calibration references: they allow us to measure how close each candidate heuristic comes to the optimum, expressed as a normalized gap. We emphasize that the search directly minimizes the heuristic's makespan; the optimal makespan from CP-SAT enters only as a constant denominator for normalization. CP-SAT does not provide structural supervision (e.g., task orderings or processor assignments) to the search procedure. Its role is to establish a quality ceiling against which candidate solutions are evaluated, enabling meaningful comparison across DAGs of different sizes and structures.

  Because the search space is low-dimensional (five weights), the procedure converges reliably within a modest evaluation budget. Once weights have been recovered across many graphs, scheduling becomes a structured prediction problem: given a new DAG, its structural characteristics determine an effective scheduling rule, connecting near-optimal quality with fast heuristic inference.

  Fig.~\ref{fig:mosaic_framework} depicts the framework. A corpus of synthetic and real benchmark DAGs is solved offline using CP-SAT to obtain optimal makespans, which provide calibration references for the staged search to recover interpretable priority weights $\theta^*$. At deployment time, the learned $\theta$ is applied within a list scheduler to produce constraint-satisfying schedules in milliseconds without invoking the solver again, decoupling the cost of exact optimization (offline, one-time) from the speed of scheduling (online, per-DAG).

\subsection{GPU Memory Hierarchy and Scheduling Implications}

A central motivation for MoSAIC-X is that scheduling decisions on modern GPUs cannot be divorced from the memory hierarchy. Understanding this hierarchy is essential to appreciate why architecture-aware scheduling matters and how MoSAIC-X exploits it. We describe each level from fastest to slowest, using the NVIDIA RTX 500 Ada Generation GPU as a concrete reference.

\paragraph{Streaming Multiprocessors (SMs).}
The RTX 500 Ada contains 16 Streaming Multiprocessors, each capable of executing a CUDA thread block independently. Each SM has its own register file and local on-chip memory. When MoSAIC-X assigns a tiled operation to an SM, the entire computation for that tile (a sub-block of a matrix multiply or element-wise operation) runs on that SM. The scheduler decides \emph{which} tile runs on \emph{which} SM and \emph{when}, subject to data dependencies between tiles.

\paragraph{Registers.}
Registers are the fastest storage on the GPU, private to each thread and accessed in a single clock cycle. During a tiled matrix multiply, the input operands and partial accumulations for a single thread's work are held in registers. The register file size per SM limits how many threads can run concurrently (occupancy). MoSAIC-X does not directly schedule at the register level, but the tile size it selects indirectly determines register pressure: larger tiles require more registers per thread, potentially reducing occupancy.

\paragraph{Shared memory (L1 / SMEM).}
Each SM has 48\,KB of on-chip shared memory that is explicitly managed by the programmer and shared across all threads in a block. This is the critical bottleneck for tiling: each tiled matrix multiply must load its input operands (two tile slices) into shared memory before computation begins. A tile size of 32 requires roughly 8\,KB of shared memory, while a tile size of 64 requires roughly 32\,KB. Both fit within the 48\,KB limit. If a tile were too large to fit, the kernel would fail to launch entirely. MoSAIC-X constrains tile sizes at DAG construction time to guarantee that every tile fits in shared memory, making this a hard architectural constraint rather than a tuning parameter.

\paragraph{L2 cache.}
The RTX 500 Ada has a 16\,MB unified L2 cache shared across all SMs. When a tile finishes execution on one SM and its output is needed by a downstream tile on another SM, the data travels through global memory but may remain resident in L2 if accessed soon enough. This creates a scheduling opportunity: if the scheduler places a consumer tile soon after the producer, the intermediate data is likely still in L2, reducing access latency compared to a full global memory fetch. MoSAIC-X's locality-aware extension (MoSAIC-X-L2) models L2 residency during scheduling. We acknowledge that because the L2 cache is shared across all SMs, co-locating tasks on the same SM does not guarantee a cache hit; temporal proximity of access matters more than SM identity. Our simulation models this by tracking a global FIFO eviction order across the aggregate L2 budget rather than partitioning per SM. Each task's input data is checked against the simulated L2 state; entries are evicted when the total working set exceeds the 16\,MB capacity.

\paragraph{Global memory (DRAM).}
Global memory (8\,GB GDDR6 on the RTX 500 Ada) is the slowest tier, accessed at 128\,GB/s bandwidth. All inter-tile data transfers that miss in L2 must go through global memory. At the Small and 1X matmul scales (defined in Section~\ref{sec:experiments}), the total working set fits within L2, so nearly all accesses hit the cache. At the 2X scale, the working set exceeds 16\,MB and spills to global memory, making scheduling order critical for performance.

\paragraph{Why this matters for scheduling.}
The memory hierarchy creates a spectrum of data access costs that depend on \emph{scheduling decisions}. Two schedules with identical makespan under a flat-memory model can differ on real hardware because one reuses L2-cached data while the other forces global memory round-trips. MoSAIC-X models this hierarchy during task dispatch, connecting graph-level priority decisions with architecture-level memory behavior.

\subsection{Detailed Methodology}

  MoSAIC-X operationalizes the above formulation through a sequence of five stages that construct calibration references from exact schedules, search for compact heuristic rules, and enable generalization to unseen workloads.

  \paragraph{Stage 1: Benchmark construction and tiling.}
  A structurally diverse corpus of DAGs is required so that the learned scheduling behavior is not tied to a single dependency pattern. The framework includes both synthetic and real application-derived task graphs. Synthetic DAGs are drawn from the Standard Task Graph set (STG)~\cite{ref14} and TGFF generator~\cite{ref15}, which provide controlled variation in topology including random sparse, layered, fork-join, and series-parallel structures. Real benchmark DAGs from PolyBench~\cite{ref11} and MachSuite~\cite{ref12} complement this by introducing dependency structures from practical computation kernels (e.g., GEMM, CONV2D, FFT, and AES).

For neural network workloads, matrix operations are decomposed into tiled sub-operations, where each tile computes a fixed-size sub-block of a matrix multiply or element-wise operation. The tile size is constrained by the GPU's shared memory: each tile's input operands must fit within 48\,KB. Each tile becomes a node in the task DAG, and data dependencies between tiles form directed edges with communication weights proportional to the tile output size. We define three matmul scales for a two-layer MLP (input projection, ReLU activation, output projection) with forward, backward, and weight-update passes: \emph{Small} ($32{\times}512{\times}256$, tile size 32, producing 588 simulation tasks), \emph{1X} ($64{\times}1024{\times}512$, tile size 64, also 588 simulation tasks), and \emph{2X} ($128{\times}2048{\times}1024$, tile size 64, producing 3{,}333 simulation tasks). Transformer training DAGs~\cite{ref27} are constructed similarly by tiling multi-head attention and feed-forward layers, with DAG sizes ranging from 78 to 1{,}024 tasks depending on the model configuration.

  \paragraph{Stage 2: Structural feature extraction.}
  Each task is described by five features that capture both its computational cost and its position within the DAG: (1)~\emph{upward rank}~\cite{ref01}, which estimates the longest weighted path from the task to any exit node; (2)~\emph{depth}, the longest path from any entry node; (3)~\emph{fanout}, the number of immediate successors; (4)~\emph{indegree}, the number of immediate predecessors; and (5)~\emph{communication cost}, the largest outgoing data transfer normalized by memory bandwidth. Together, fanout and indegree capture local connectivity patterns: high fanout identifies fork regions where parallelism fans out, while high indegree identifies join points where parallel branches converge. We refer to this as ``structural sensitivity'' rather than explicit motif detection; no subgraph enumeration is performed. Because all five features are defined over structural properties common to any task graph, the learned weights transfer naturally across DAG families without recomputation.

\begin{figure}[!htbp]
\centering
\begin{tcolorbox}[
    colback=gray!5,
    colframe=black,
    boxrule=0.4pt,
    arc=6pt,
    width=\columnwidth,
    left=2pt,right=2pt,top=4pt,bottom=4pt
]
\centering
\resizebox{0.95\columnwidth}{!}{%
\begin{tikzpicture}[
    every node/.style={font=\scriptsize},
    motifnode/.style={circle, draw=black, thick, minimum size=6mm, inner sep=0pt},
    forkjoin/.style={motifnode, fill=violet!25},
    fanout/.style={motifnode, fill=green!25},
    fanin/.style={motifnode, fill=orange!30},
    titlebox/.style={font=\bfseries\scriptsize},
    >=Latex,
    thick
]

\node[titlebox] at (0,1.8) {Fork-Join};

\node[forkjoin] (f1) at (0,0.9) {1};
\node[forkjoin] (f2) at (-0.8,0) {2};
\node[forkjoin] (f3) at (0,0) {3};
\node[forkjoin] (f4) at (0.8,0) {4};
\node[forkjoin] (f0) at (0,-0.9) {0};

\draw[->] (f1) -- (f2);
\draw[->] (f1) -- (f3);
\draw[->] (f1) -- (f4);
\draw[->] (f2) -- (f0);
\draw[->] (f3) -- (f0);
\draw[->] (f4) -- (f0);

\node[titlebox] at (3.8,1.8) {Fan-Out};

\node[fanout] (o4) at (3.8,0.9) {4};
\node[fanout] (o0) at (2.6,0) {0};
\node[fanout] (o1) at (3.4,0) {1};
\node[fanout] (o2) at (4.2,0) {2};
\node[fanout] (o3) at (5.0,0) {3};

\draw[->] (o4) -- (o0);
\draw[->] (o4) -- (o1);
\draw[->] (o4) -- (o2);
\draw[->] (o4) -- (o3);

\node[titlebox] at (7.7,1.8) {Fan-In};

\node[fanin] (i1) at (6.7,0.9) {1};
\node[fanin] (i2) at (7.4,0.9) {2};
\node[fanin] (i3) at (8.1,0.9) {3};
\node[fanin] (i4) at (7.4,0) {4};
\node[fanin] (i0) at (7.4,-0.9) {0};

\draw[->] (i1) -- (i4);
\draw[->] (i2) -- (i4);
\draw[->] (i3) -- (i4);
\draw[->] (i4) -- (i0);

\end{tikzpicture}
}
\end{tcolorbox}

\caption{Representative recurring structural patterns in task DAGs: fork-join (synchronization barrier), fan-out (parallel branch point), and fan-in (merge point). The learned weights on fanout and indegree features capture sensitivity to these patterns without explicit subgraph enumeration.}
\label{fig:motif_classes}
\end{figure}
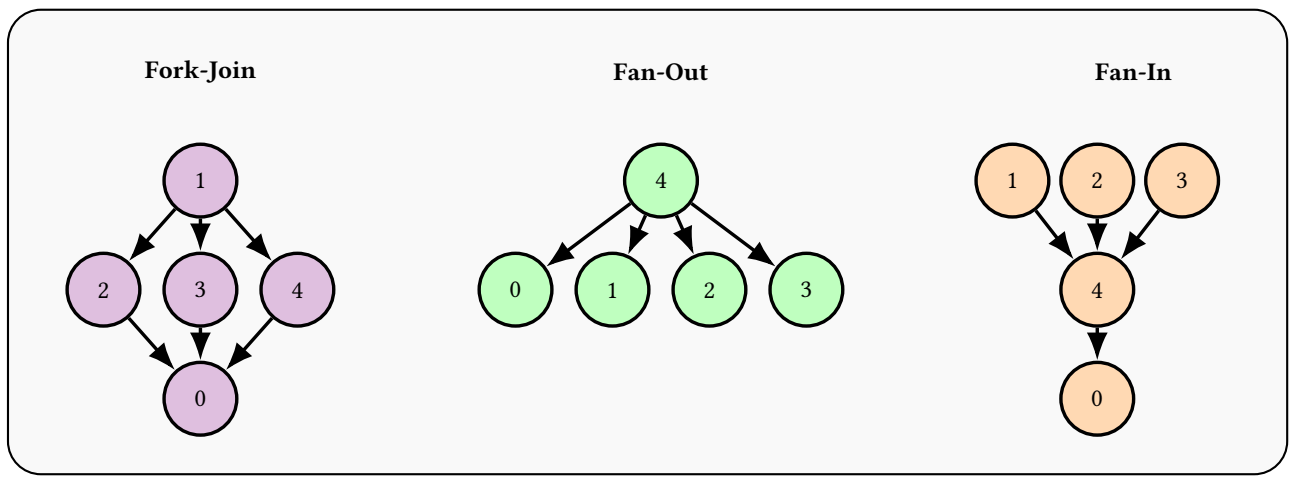

Fig.~\ref{fig:motif_classes} illustrates representative recurring structural patterns. The learned weights on fanout and indegree control how aggressively the scheduler prioritizes or defers tasks at structurally distinct positions. Unlike explicit motif enumeration via subgraph mining~\cite{ref26}, this implicit encoding avoids the computational cost of motif detection while still capturing the scheduling-relevant aspects of local graph structure. We note that the contribution of individual features has not been isolated through ablation; the five features were selected based on prior scheduling literature~\cite{ref01, ref03, ref22} and their joint effect is what the search optimizes.

  \paragraph{Stage 3: Optimal makespan computation.}
  Each DAG is formulated as a constraint programming problem and solved using the CP-SAT solver~\cite{ref07}. The solver assigns tasks to processors, enforces precedence and communication constraints, and minimizes overall execution time. For small-to-medium instances (up to roughly a thousand tasks), CP-SAT returns provably optimal solutions; for the 960-task Transformer-Base instance, a 30-minute time limit is imposed and the best feasible solution serves as the reference. The optimal makespan for each DAG provides the calibration signal for the subsequent search stage. We stress that only the scalar makespan value is used; the solver's internal task ordering and processor assignments are not passed to the search procedure.

\paragraph{Stage 4: Priority weight search via staged stochastic optimization.}
  For a given set of weights $\theta \in \mathbb{R}^5$, list scheduling assigns tasks in decreasing order of priority $H(v;\theta) = \theta^\top \varphi(v)$ to the processor that yields the earliest finish time, respecting precedence and communication constraints. The objective is to minimize the heuristic's makespan $C_{\max}^{H(\theta)}$. The normalized gap $(C_{\max}^{H(\theta)} - C_{\max}^*) / C_{\max}^*$ is used to compare quality across DAGs of different sizes, but minimizing this gap is mathematically equivalent to minimizing $C_{\max}^{H(\theta)}$ since $C_{\max}^*$ is a positive constant for each DAG.

  The search proceeds in three phases: (1)~\emph{broad exploration}, drawing 1{,}000 weight vectors from a wide Gaussian distribution and evaluating each by running the list scheduler; (2)~\emph{targeted refinement}, taking the top candidates from phase 1 and perturbing them with 500 smaller-variance samples; (3)~\emph{fine-tuning}, applying 200 local perturbations around the best solution found so far. The total budget is approximately 1{,}700 candidate evaluations per DAG, each taking under a millisecond for moderate-sized graphs.

  We note that this procedure is a staged stochastic optimization~\cite{ref08} rather than surrogate-based Bayesian Optimization, as it does not maintain a probabilistic model of the objective. The 1{,}700-evaluation budget gives MoSAIC-X more computation than single-shot baselines like HEFT (expressible as $\theta = (1, 0, 0, 0, 0)$ in our framework). Comparison against alternative search methods under the same budget is left to future work.

  Across the benchmark corpus, the recovered weights consistently assign the largest importance to upward rank, confirming it as the dominant scheduling signal. The remaining weights adapt to graph structure: communication-heavy DAGs emphasize communication cost, hub-dominated structures penalize premature fan-out expansion, and merge-heavy regions boost the priority of synchronization-critical join nodes.

  \paragraph{Stage 5: Transfer, memory hierarchy, and deployment.}
  Once weights have been recovered on source DAGs, they can be applied zero-shot to unseen targets because the features are defined over structural properties intrinsic to any task graph. In practice, weights learned from a small 50-task graph can directly schedule a much larger Transformer DAG with nearly a thousand tasks. The knowledge base stores $\{(G_i, \varphi_i, \theta_i^*)\}$ triples and selects transfer weights by matching the source DAG that is most structurally similar to the target (based on average depth, density, and CCR). As shown in Section~\ref{sec:experiments}, same-family transfer retains meaningful benefit, while cross-family transfer yields results comparable to HEFT.

  For locality-sensitive workloads, MoSAIC-X extends the scheduler with memory-hierarchy awareness. The scheduler tracks which tile outputs are likely still resident in L2 cache (modeled as a global FIFO buffer). Data recently produced by a predecessor is assumed cached, reducing access time compared to global memory. The learned weights exploit this by prioritizing tasks that maximize data reuse.

\begin{figure}[!htbp]
  \centering
  \includegraphics[width=\columnwidth]{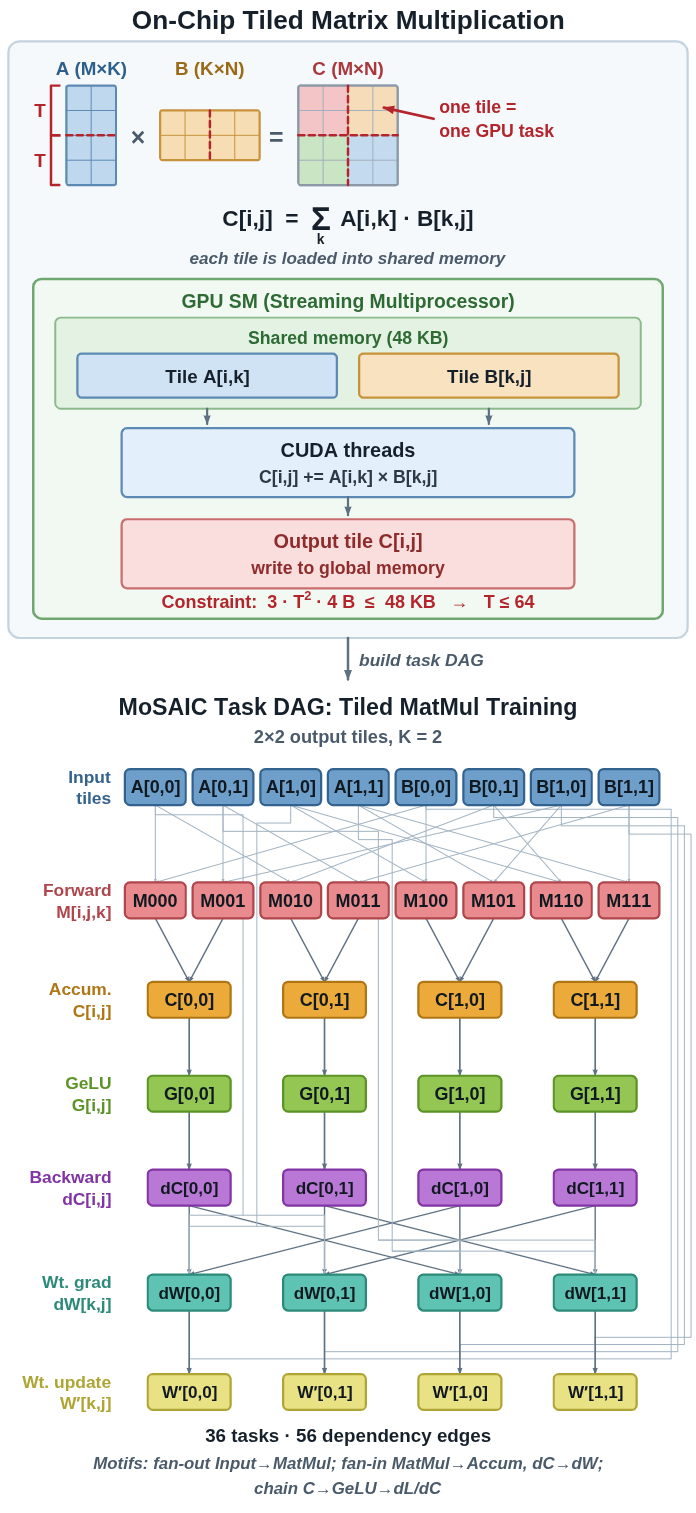}
  \caption{On-chip tiled matrix multiplication and the resulting MoSAIC-X task DAG for a $2{\times}2$ tiled MatMul training iteration (36 tasks, 56 dependency edges). Top: each tile pair is loaded into shared memory and multiplied by a CUDA kernel on a single SM. Bottom: the full forward, backward, and weight update DAG exposing fork, fan-out, fan-in, and chain patterns.}
  \label{fig:case_study_dag}
\end{figure}

\textbf{Case Study: Tiled MatMul Training DAG.}
  Figure~\ref{fig:case_study_dag} illustrates MoSAIC-X on a minimal example: a matrix multiply tiled into a small grid of blocks that each fit in shared memory. Input tile pairs are loaded and multiplied by CUDA kernels, producing independent MatMul tasks that fan out from the input loads. Partial products are then reduced (fan-in) into accumulation tasks, each followed by activation, backward gradient, weight gradient, and weight update stages, forming a chain per output tile. Cross-tile edges arise at the weight-gradient stage, where each update depends on gradients and inputs from different rows and columns. The resulting DAG exposes structural patterns (fan-out, fan-in, and chain) that the learned priority weights can exploit to schedule critical-path tasks first and co-locate producer-consumer tiles for data reuse.

 \section{Experimental Setup and Evaluation}
  \label{sec:experiments}

The evaluation measured how effectively MoSAIC-X recovered near-optimal scheduling behavior and transferred it across diverse DAG families and matrix scales. All simulation experiments targeted the NVIDIA RTX 500 Ada Generation GPU architecture with 16 streaming multiprocessors, 48\,KB shared memory per block, 16\,MB L2 cache, and 128\,GB/s bandwidth, unless otherwise noted. The simulation used 4 abstract processors (PEs) for most experiments; each PE modeled an SM group. Tables~\ref{tab:ccr} and~\ref{tab:locality} varied the number of PEs as noted in their captions. The learned model consisted of just five scalar weights used within a standard list scheduler.

  \subsection{Benchmarks}

  The corpus comprised 20 DAGs spanning synthetic and real application kernels. Synthetic DAGs included STG families~\cite{ref14} (such as FFT, Gauss, Laplace, and Montage) and TGFF-generated~\cite{ref15} heterogeneous SoC graphs. Real benchmark DAGs were derived from PolyBench~\cite{ref11} and MachSuite~\cite{ref12} kernels including GEMM, AES, CONV2D, and others. Transformer training DAGs~\cite{ref27} were constructed at four scales by tiling multi-head attention and feed-forward layers (Tiny: 78 tasks, Small: 456 tasks, Medium: 960 tasks, Base: 1{,}024 tasks). A matmul case study further decomposed two-layer MLP training into tiled operations at three matrix scales (Small, 1X, 2X), producing simulation DAGs ranging from 588 to 3{,}333 tasks.

  \textbf{Note on task counts.} The simulation DAG and the hardware execution DAG differed in task count for the same scale label because the simulation included fine-grained communication and synchronization nodes that were fused into larger CUDA kernels during actual GPU execution. For example, the Small matmul configuration had 588 tasks in simulation (Table~\ref{tab:tiling}) but 168 CUDA kernel invocations on hardware (Table~\ref{tab:gpu_valid}); similarly, 1X had 588 simulation tasks but 901 kernel invocations due to different granularity in the backward pass decomposition. This distinction is important for interpreting the correspondence between simulated and measured results.

\subsection{Tiling and Memory Hierarchy}

  Neural network training was decomposed into tiled matrix operations, where each tile computed a sub-block of a matrix multiply or element-wise operation. As described in Section~\ref{sec:method}, tile size was constrained at multiple levels of the GPU memory hierarchy. Table~\ref{tab:tiling} reports how the matmul case study interacted with each hierarchy level across three matrix scales.

  At the Small scale, each tile used only 8\,KB of shared memory (well within the 48\,KB limit), and the total working set of 0.57\,MB fit easily in the 16\,MB L2 cache. At the 1X scale, shared memory usage rose to 32\,KB per tile (still within limits), and the 2.27\,MB working set remained L2-resident. At the 2X scale, the working set exceeded L2 capacity entirely, forcing frequent global memory accesses. This is where scheduling order mattered most: by co-locating producer and consumer tiles temporally, MoSAIC-X-L2 kept intermediate data in L2 longer and avoided costly DRAM round-trips.

  Table~\ref{tab:matmul_sched} compared scheduling quality on these matmul DAGs. All heuristics produced nearly identical makespans, which we treated as ties within simulation precision. This is the expected outcome for highly regular graphs with symmetric fan-in/fan-out structure: when all tasks have similar weights and uniform connectivity, any reasonable priority ordering produces a near-optimal schedule~\cite{ref23}. The value of learned weights emerges on \emph{irregular} graphs with heterogeneous structure, as shown in the following subsections.

  \begin{table}[!htbp]
  \centering
  \caption{Matmul tiling: matrix scaling and memory hierarchy on RTX 500 Ada (4 PEs). All tile sizes fit within the 48\,KB shared memory limit.}
  \label{tab:tiling}
  \setlength{\tabcolsep}{3pt}
  \renewcommand{\arraystretch}{1.1}
  \footnotesize
  \fbox{%
  \begin{tabular}{@{\hspace{4pt}}lrrrrrrr@{\hspace{4pt}}}
  \toprule
  \textbf{Scale} & \textbf{Dims} & \textbf{Tile} & \textbf{Tasks} & \textbf{Smem} & \textbf{Occ.} & \textbf{WS} & \textbf{L2} \\
   & & & & \textbf{(KB)} & \textbf{(\%)} & \textbf{(MB)} & \textbf{fit?} \\
  \midrule
  Small & $32{\times}512{\times}256$ & 32 &  588 &  8 & 66.7 & 0.57 & Yes \\
  1X    & $64{\times}1024{\times}512$ & 64 &  588 & 32 & 66.7 & 2.27 & Yes \\
  2X    & $128{\times}2048{\times}1024$ & 64 & 3333 & 32 & 66.7 & --- & No \\
  \bottomrule
  \end{tabular}%
  }
  \end{table}

\begin{table}[!htbp]
  \centering
  \caption{Matmul scheduling results (4 PEs). Makespan in $\mu$s. Differences below 0.01\% ($\dagger$) are treated as ties due to simulation precision.}
  \label{tab:matmul_sched}
  \setlength{\tabcolsep}{3pt}
  \renewcommand{\arraystretch}{1.1}
  \footnotesize
  \fbox{%
  \begin{tabular}{@{\hspace{4pt}}lrrrrrr@{\hspace{4pt}}}
  \toprule
  \textbf{Scale} & \textbf{HEFT} & \textbf{CPOP} & \textbf{FDS} & \textbf{DLS} & \textbf{MoSAIC-X} & \textbf{Eff.} \\
  \midrule
  Small & 10079.5 & 10105.1 & \textbf{10078.5} & 10079.5 & 10078.7$^\dagger$ & 1.00 \\
  1X    & 79217.8 & 79313.8 & \textbf{79215.8} & 79217.8 & 79216.9$^\dagger$ & 1.00 \\
  2X    & 629023  & 629157  & \textbf{629009}  & 629023  & 629009$^\dagger$  & 1.00 \\
  \bottomrule
  \end{tabular}%
  }
  \end{table}

  \subsection{Results on Community Benchmarks}

  Table~\ref{tab:real_perf} reported simulated makespan (in $\mu$s) on eight PolyBench and MachSuite kernels using 4 PEs, with CP-SAT~\cite{ref07} providing the optimal reference (seven of eight proven optimal within the time limit). MoSAIC-X was compared against HEFT~\cite{ref01}, CPOP~\cite{ref01}, FDS~\cite{ref04}, and DLS~\cite{ref03}. MoSAIC-X achieved the best or tied-best makespan on seven of eight benchmarks. On GEMM, the largest kernel (63 tasks), MoSAIC-X achieved 164.20\,$\mu$s, within 0.1\% of the CP-SAT optimum (164.04\,$\mu$s) and 0.89\% faster than HEFT (165.68\,$\mu$s). STENCIL was the only case where FDS narrowly outperformed MoSAIC-X (6.47 vs.\ 6.48\,$\mu$s, a difference of 0.15\%). These results demonstrate that the learned priority weights consistently recover near-optimal schedules across diverse kernel structures.

  \begin{table}[!htbp]
  \centering
  \caption{Simulated makespan ($\mu$s) on PolyBench and MachSuite benchmarks (4 PEs). Bold indicates best heuristic result; CP-SAT is shown separately as the reference optimum.}
  \label{tab:real_perf}
  \setlength{\tabcolsep}{3pt}
  \renewcommand{\arraystretch}{1.1}
  \footnotesize
  \fbox{%
  \begin{tabular}{@{\hspace{4pt}}lr|rrrrr@{\hspace{4pt}}}
  \toprule
  \textbf{Kernel} & \textbf{Tasks} & \textbf{CP-SAT} & \textbf{HEFT} & \textbf{CPOP} & \textbf{FDS} & \textbf{DLS} \\
  \midrule
  GEMM    &  63 & 164.04 & 165.68 & 169.90 & 164.26 & 165.68 \\
  SYRK    &  39 & 107.68 & 107.86 & 114.66 & 109.30 & 107.86 \\
  FFT     & 112 &   5.60 &   5.60 &   6.20 &   5.60 &   5.60 \\
  AES     & 131 &  36.57 &  36.86 &  38.91 &  39.32 &  36.86 \\
  ATAX    &  60 &   3.79 &   3.88 &   4.29 &   3.89 &   3.88 \\
  BICG    &  64 &   3.97 &   4.09 &   4.19 &   4.13 &   4.09 \\
  CONV2D  & 144 &  14.12 &  14.47 &  14.52 &  14.42 &  14.47 \\
  STENCIL &  64 &   6.37 &   6.48 &   6.96 &   \textbf{6.47} &   6.48 \\
  \bottomrule
  \end{tabular}%
  }

  \vspace{2pt}
  \fbox{%
  \begin{tabular}{@{\hspace{4pt}}lr|r@{\hspace{4pt}}}
  \toprule
  \textbf{Kernel} & \textbf{Tasks} & \textbf{MoSAIC-X} \\
  \midrule
  GEMM    &  63 & \textbf{164.20} \\
  SYRK    &  39 & \textbf{107.77} \\
  FFT     & 112 & \textbf{5.60} \\
  AES     & 131 & \textbf{36.86} \\
  ATAX    &  60 & \textbf{3.87} \\
  BICG    &  64 & \textbf{4.09} \\
  CONV2D  & 144 & \textbf{14.38} \\
  STENCIL &  64 & 6.48 \\
  \bottomrule
  \end{tabular}%
  }
  \end{table}

  \subsection{Near-Optimality Verification \& Architecture Variation}

  Table~\ref{tab:cpsat_opt} verified MoSAIC-X against proven CP-SAT optima on Transformer DAGs and synthetic topologies using 4 PEs. All instances stayed within half a percent of the proven optimum. On the 960-task Transformer-Base, the largest instance where CP-SAT could certify optimality within a 30-minute time limit, MoSAIC-X achieved a gap of just 0.01\%, outperforming HEFT's 0.03\% gap. This confirms that the five learned weights recover scheduling quality comparable to exhaustive constraint solving, at a fraction of the computational cost.

  Table~\ref{tab:ccr} presented a CCR sweep on a large Transformer DAG (4 PEs), where CCR is the ratio of average communication cost to average computation cost. This experiment revealed where MoSAIC-X's adaptive weighting provided the greatest advantage. In communication-dominated regimes (CCR $\geq$ 5.0), MoSAIC-X reduced makespan by up to 11.74\% relative to HEFT, a substantial improvement that fixed heuristics cannot achieve because they do not reweight features based on the communication-computation balance. At CCR = 1.0, FDS (244.00\,$\mu$s) outperformed both MoSAIC-X (254.64\,$\mu$s) and HEFT (266.40\,$\mu$s); at CCR = 0.1 and 10.0, MoSAIC-X matched FDS. This pattern suggests that MoSAIC-X's learned weights successfully emphasize communication cost when it dominates the schedule, precisely the regime where architecture-aware scheduling matters most.

  \begin{table}[!htbp]
  \centering
  \caption{MoSAIC-X gap from verified CP-SAT optimal makespans ($\mu$s), 4 PEs. HEFT gap shown for comparison.}
  \label{tab:cpsat_opt}
  \setlength{\tabcolsep}{3pt}
  \renewcommand{\arraystretch}{1.1}
  \footnotesize
  \fbox{%
  \begin{tabular}{@{\hspace{4pt}}lrrrrr@{\hspace{4pt}}}
  \toprule
  \textbf{Instance} & \textbf{Tasks} & \textbf{CP-SAT} & \textbf{HEFT Gap} & \textbf{MoSAIC-X} & \textbf{Gap\,(\%)} \\
  \midrule
  Trans-Small  &  28 & 10.83 & 0.74 & 10.88 & 0.49 \\
  Trans-Medium &  76 & 141.82 & 0.11 & 141.87 & 0.04 \\
  Trans-Base   & 960 & 2490.32 & 0.03 & 2490.58 & 0.01 \\
  Chain-20           &  20 & 102.30 & 0.17 & 102.39 & 0.08 \\
  Diamond-30         &  30 & 106.58 & 0.15 & 106.67 & 0.08 \\
  ForkJoin-40        &  40 & 155.10 & 0.50 & 155.49 & 0.25 \\
  Random-50          &  50 & 125.21 & 0.22 & 125.35 & 0.11 \\
  \bottomrule
  \end{tabular}%
  }
  \end{table}

  \begin{table}[!htbp]
  \centering
  \caption{CCR sweep on Transformer DAG (4 PEs). Makespan in $\mu$s. Gap is MoSAIC-X vs.\ HEFT (\%); negative means MoSAIC-X is better. Note: at CCR = 1.0, FDS outperforms MoSAIC-X.}
  \label{tab:ccr}
  \setlength{\tabcolsep}{5pt}
  \renewcommand{\arraystretch}{1.1}
  \footnotesize
  \fbox{%
  \begin{tabular}{@{\hspace{4pt}}lrrrr@{\hspace{4pt}}}
  \toprule
  \textbf{CCR} & \textbf{HEFT} & \textbf{Best Classical} & \textbf{MoSAIC-X} & \textbf{Gap\,(\%)} \\
  \midrule
  0.1  & 247.67 & 241.28 (FDS) & 241.28 &  $-$2.58 \\
  1.0  & 266.40 & 244.00 (FDS) & 254.64 &  $-$4.42 \\
  5.0  & 408.59 & 393.59 (FDS) & 360.53 & $-$11.74 \\
  10.0 & 621.34 & 572.83 (FDS) & 572.83 &  $-$7.82 \\
  \bottomrule
  \end{tabular}%
  }
  \end{table}

  \subsection{Locality-Aware Scheduling}

  MoSAIC-X extended list scheduling with L2-cache-aware runtime adjustment (Table~\ref{tab:locality}). During scheduling, the simulator tracked which tile outputs were likely still resident in L2 cache (modeled as a global FIFO buffer with 16\,MB capacity). When a task's input had been recently produced and had not been evicted, memory access was modeled as faster than a full global memory fetch. On the Tiny configuration (78 tasks), MoSAIC-X-L2 improved 4.70\% over the locality-aware HEFT variant, demonstrating that cache-aware priority adjustment can meaningfully reduce makespan when the working set is small enough to create eviction pressure. On larger configurations, the 16\,MB L2 cache accommodated most working sets, naturally reducing locality pressure, an expected outcome that validates the memory hierarchy model rather than indicating a limitation of the approach. Hardware validation of cache behavior with profiling tools such as NVIDIA Nsight Compute remains future work.

  \begin{table}[!htbp]
  \centering
  \caption{Locality-aware scheduling on Transformer DAGs (4 PEs). Makespan in $\mu$s. Gap is MoSAIC-X-L2 vs.\ L2-HEFT (\%).}
  \label{tab:locality}
  \setlength{\tabcolsep}{4pt}
  \renewcommand{\arraystretch}{1.1}
  \footnotesize
  \fbox{%
  \begin{tabular}{@{\hspace{4pt}}lrrrrr@{\hspace{4pt}}}
  \toprule
  \textbf{Config} & \textbf{Tasks} & \textbf{HEFT} & \textbf{L2-HEFT} & \textbf{MoSAIC-X-L2} & \textbf{Gap\,(\%)} \\
  \midrule
  Tiny   &   78 &  5.09 &  5.20 &  4.95 & $-$4.70 \\
  Small  &  456 & 72.04 & 72.72 & 72.69 & $-$0.04 \\
  Medium &  960 & 1245.34 & 1251.72 & 1250.77 & $-$0.08 \\
  Base   & 1024 & 1255.23 & 1260.89 & 1260.22 & $-$0.05 \\
  \bottomrule
  \end{tabular}%
  }
  \end{table}

\subsection{Real GPU Validation and Comparison with DeepSoCS}

  To validate that simulation-predicted improvements translated to real hardware, both HEFT and MoSAIC-X task orderings were executed on the RTX 500 Ada using PyTorch CUDA kernels (Table~\ref{tab:gpu_valid}). On the 1X configuration (901 kernel invocations), MoSAIC-X achieved a $1.015\times$ speedup over HEFT, confirming that the learned task ordering produced measurable performance gains on real silicon. The Small configuration (168 kernel invocations) showed no significant difference ($0.992\times$, with overlapping standard deviations), consistent with the simulation prediction that regular small DAGs offer minimal scheduling headroom. The hardware execution mapped each scheduled task to a CUDA kernel launch; the CUDA runtime's SM assignment was not directly controlled, so these results reflect the benefit of task \emph{ordering} rather than processor \emph{binding}.

  Table~\ref{tab:deepsocs} compared MoSAIC-X against DeepSoCS~\cite{ref09}, a neural scheduler using deep reinforcement learning with graph neural network embeddings. We reimplemented DeepSoCS with a two-layer GNN and a fully connected policy head totaling 417 parameters, trained for 500 episodes per graph. At 200 tasks, MoSAIC-X achieved a 2.05\% improvement over HEFT while DeepSoCS degraded by 48.65\%, illustrating a key advantage of the compact parametric approach: MoSAIC-X's five weights generalize naturally to larger graphs because the underlying features (upward rank, depth, fanout, indegree, communication cost) are scale-invariant, whereas neural policies may require additional training data to handle increased graph complexity. We note that additional hyperparameter tuning and longer training for DeepSoCS could narrow this gap.

  \begin{table}[!htbp]
  \centering
  \caption{GPU execution on RTX 500 Ada (PyTorch CUDA kernels). Measured over 50 independent
  runs; $\pm$ denotes one standard deviation. All other tables report
  \emph{simulated} makespans, not measured GPU times. Task counts differ from simulation tables because CUDA kernel granularity differs from the simulation DAG decomposition (see Section~\ref{sec:experiments}).}
  \label{tab:gpu_valid}
  \setlength{\tabcolsep}{4pt}
  \renewcommand{\arraystretch}{1.1}
  \footnotesize
  \fbox{%
  \begin{tabular}{@{\hspace{4pt}}lrrrr@{\hspace{4pt}}}
  \toprule
  \textbf{Config} & \textbf{Tasks} & \textbf{HEFT ($\mu$s)} & \textbf{MoSAIC-X ($\mu$s)} & \textbf{Speedup} \\
  \midrule
  Small & 168 & 2616.6 $\pm$ 67.7 & 2638.2 $\pm$ 53.3 & 0.992$\times$ \\
  1X    & 901 & 18213.2 $\pm$ 318.3 & 17942.6 $\pm$ 356.6 & 1.015$\times$ \\
  \bottomrule
  \end{tabular}%
  }
  \end{table}

  \begin{table}[!htbp]
  \centering
  \caption{MoSAIC-X vs.\ DeepSoCS (4 PEs). Simulated makespan in $\mu$s. Gap relative to HEFT (\%).}
  \label{tab:deepsocs}
  \setlength{\tabcolsep}{4pt}
  \renewcommand{\arraystretch}{1.1}
  \footnotesize
  \fbox{%
  \begin{tabular}{@{\hspace{4pt}}lrrrrr@{\hspace{4pt}}}
  \toprule
  \textbf{Tasks} & \textbf{HEFT} & \textbf{MoSAIC-X} & \textbf{DeepSoCS} & \textbf{M-Gap} & \textbf{D-Gap} \\
  \midrule
    20 & 124.37 & 124.37 & 124.37 &  0.00 &   0.00 \\
   100 & 183.84 & 183.66 & 188.65 & $-$0.10 & $+$2.62 \\
   200 & 213.34 & 208.96 & 317.12 & $-$2.05 & $+$48.65 \\
  \bottomrule
  \multicolumn{6}{@{\hspace{4pt}}l}{\footnotesize MoSAIC-X: 5 params, 1700 evals. DeepSoCS: 417 params, 500 episodes.}
  \end{tabular}%
  }
  \end{table}

  \subsection{Zero-Shot Transfer and Scalability}

  \begin{table}[!htbp]
  \centering
  \caption{Zero-shot transfer (4 PEs): gap relative to HEFT (\%). Negative means better than HEFT. Native = per-instance optimization. Ensemble = oracle best over all stored weight vectors (upper bound, not a practical strategy).}
  \label{tab:transfer}
  \setlength{\tabcolsep}{4pt}
  \renewcommand{\arraystretch}{1.1}
  \footnotesize
  \fbox{%
  \begin{tabular}{@{\hspace{4pt}}llrrr@{\hspace{4pt}}}
  \toprule
  \textbf{Target} & \textbf{Tasks} & \textbf{Transfer} & \textbf{Native} & \textbf{Ensemble} \\
  \midrule
  TGFF-200      &  200 & $-$0.17 & $-$1.31 & $+$2.03 \\
  TGFF-200-ccr5 &  200 & $-$1.22 & $-$8.24 & $+$3.52 \\
  TGFF-500      &  500 & $+$0.19 & $-$0.17 & $+$0.47 \\
  Trans-Base    & 1024 & $+$0.09 & $-$0.05 & $+$0.08 \\
  \bottomrule
  \end{tabular}%
  }
  \end{table}

  Table~\ref{tab:transfer} evaluated whether weights learned on small source DAGs transferred zero-shot to unseen targets~\cite{ref25}. Same-family transfer was effective: TGFF-200 and TGFF-200-ccr5 retained meaningful improvements over HEFT ($-0.17\%$ and $-1.22\%$ respectively), demonstrating that the learned weights capture transferable structural properties within a DAG family. Cross-family transfer to TGFF-500 and Trans-Base yielded makespans within 0.2\% of HEFT, neither a significant improvement nor a significant degradation, suggesting that cross-family transfer is neutral rather than harmful. The Native column shows that per-instance optimization consistently outperforms transfer, indicating that the best results require tuning weights to each workload. The Ensemble column reports an oracle upper bound: the best makespan achievable by evaluating all stored weight vectors at scheduling time.

\subsection{Summary and Limitations}

  MoSAIC-X achieved near-optimal schedules using five learned parameters, outperforming HEFT on the majority of per-instance benchmarks with gains of up to 12\% in communication-heavy regimes. It scaled from hundreds to thousands of tasks, delivered measurable speedup on real GPU hardware, and used orders of magnitude fewer parameters than neural alternatives like DeepSoCS~\cite{ref09}.

  Several limitations should be noted. First, the 1{,}700-evaluation search budget gave MoSAIC-X more computation than single-shot baselines like HEFT; isolating the contribution of the search budget from the method itself requires comparison against alternative search strategies under the same budget, which we leave to future work. Second, zero-shot transfer was effective within DAG families but did not consistently improve over HEFT across dissimilar topologies, indicating that per-instance optimization remains necessary for best results. Third, hardware validation covered two configurations on a single GPU; broader validation across architectures and workload sizes would strengthen the generality claims. Fourth, the five structural features were selected from prior literature without ablation; quantifying individual feature contributions would clarify which aspects of graph structure drive the scheduling improvements.

\section{Conclusion}
  This paper presented MoSAIC-X, a framework that constructs interpretable, architecture-aware scheduling heuristics for GPU-accelerated neural network training. The central finding is that a five-parameter linear priority function over structural DAG features, when calibrated against exact solver solutions, can recover near-optimal scheduling quality across diverse workloads, from standard PolyBench/MachSuite kernels to Transformer training DAGs with thousands of tasks. MoSAIC-X achieved makespans within 0.5\% of proven optima, reduced makespan by up to 12\% over HEFT on communication-heavy workloads, delivered measurable speedup on real NVIDIA RTX 500 Ada hardware, and matched or outperformed a 417-parameter neural scheduler using just five interpretable weights.

  These results suggest that the DAG scheduling problem admits far more compact solutions than current neural approaches assume: the structure needed for effective scheduling can be captured by a handful of well-chosen features rather than learned implicitly through deep networks. The recovered priority weights remain fully interpretable, enabling practitioners to understand \emph{why} a particular scheduling decision was made.

  Future work will focus on three directions: (1)~ablation studies to quantify individual feature contributions and guide feature selection for new architectures, (2)~comparison of the staged stochastic search against alternative optimization methods under equivalent budgets, and (3)~hardware profiling with cache hit rates and SM utilization metrics to validate the locality-aware scheduling model on real silicon.

\end{document}